\documentclass[aps,prl,reprint,superscriptaddress]{revtex4-2}

\usepackage{graphicx}
\usepackage{colortbl}
\usepackage{amsmath}
\usepackage{bm}
\usepackage{color}
\usepackage{soul}

\makeatletter
\def\btt#1{\texttt{\@backslashchar#1}}%
\DeclareRobustCommand\bblash{\btt{\@backslashchar}}%
\makeatother

\begin{document}


\title{Thermally quenched metastable phase in the Ising model with competing interactions} 



\author{Hiroshi Oike}
\email{OIKE.Hiroshi@nims.go.jp}
\affiliation{PRESTO, Japan Science and Technology Agency (JST), Saitama 332-0012, Japan}
\affiliation{Research Center for Materials Nanoarchitechtonics (MANA), National Institute for Materials Science (NIMS), Ibaraki 305-0047, Japan}
\affiliation{Department of Applied Physics and Quantum-Phase Electronics Centre (QPEC), The University of Tokyo, Tokyo 113-8656, Japan}

\author{Hidemaro Suwa}
\affiliation{Department of Physics, The University of Tokyo, Tokyo 113-0033, Japan}

\author{Yasunori Takahashi}
\affiliation{Department of Applied Physics and Quantum-Phase Electronics Centre (QPEC), The University of Tokyo, Tokyo 113-8656, Japan}

\author{Fumitaka Kagawa}
\affiliation{Department of Applied Physics and Quantum-Phase Electronics Centre (QPEC), The University of Tokyo, Tokyo 113-8656, Japan}
\affiliation{Department of Physics, Tokyo Institute of Technology, Tokyo 152-8551, Japan}


\date{\today}

\begin{abstract}
Thermal quenching has been used to find metastable materials such as hard steels and metallic glasses. More recently, quenching-based phase control has been applied to correlated electron systems that exhibit metal--insulator, magnetic or superconducting transitions. Despite the discovery of metastable electronic phases, however, how metastability is achieved through the degrees of freedom, which can vary even at low temperatures such as those of an electron, is unclear. Here, we show a thermally quenched metastable phase in the Ising model without conservation of magnetization by Monte Carlo simulations. When multiple types of interactions that stabilize different long-range orders are introduced, the ordering kinetics divergently slow toward low temperatures, meaning that the system will reach a low temperature without ordering if the cooling rate is high enough. Quantitative analysis of the divergent behavior suggests that the energy barrier for eliminating the local structure of competing orders is the origin of this metastability. Thus, the present simulations show that competing interactions play a key role in realizing metastability.
\end{abstract}


\maketitle

\section{Introduction}

The thermodynamically most stable phase of a material is uniquely determined when thermodynamic parameters such as the temperature, pressure, and chemical composition are fixed. In contrast, multiple metastable phases may exist for a given thermodynamic parameter. For example, although an aggregate of carbon atoms has graphite as the most stable phase under ambient conditions, several allotropes, such as diamond and lonsdaleite, exist as metastable phases \cite{derjaguin1977nature}. Using metastable phases as an exploration space, solid-state chemistry and metallurgy have been developed \cite{sun2016thermodynamic, sun2017thermodynamic, martinolich2017toward, bykov2019high, ito2023stability}. A metastable phase is often achieved via thermal quenching, as exemplified by the glassy state that occurs when a liquid is cooled so rapidly that crystallization is kinetically avoided \cite{uhlmann1972kinetic, greer1995metallic, angell1995formation, debenedetti2001supercooled}. More recently, quenching-based phase control has been applied to correlated electron materials, realizing metastable electronic phases such as charge glasses \cite{kagawa2013charge, oike2015phase, sasaki2017crystallization}, charge density waves \cite{yoshida2014controlling}, magnetic skyrmions \cite{oike2016interplay, berruto2018laser, yu2018aggregation, birch2019increased}, orbital disordered phases \cite{katsufuji2020nucleation}, ferromagnetic phases \cite{matsuura2021kinetic, matsuura2023thermodynamic}, and superconductivity \cite{oike2018kinetic}. Furthermore, the metastable phase of correlated electrons is also realized through intense excitation of electrons using pulsed lasers \cite{stojchevska2014ultrafast, vaskivskyi2015controlling}. Thus, metastability is becoming a widely researched topic in materials science.

Along with the development of methods for making glasses, their design principles have also been clarified. Supercooled liquids tend to have a high glass forming ability near a eutectic point (Fig.~1(a)), as exemplified by molten alloys \cite{greer1995metallic}, pressurized silicon \cite{molinero2006tuning}, and salt-added water \cite{kobayashi2011possible}. The microscopic mechanism underlying this behavior has been investigated \cite{waseda1975structure, weber1985local, tanaka2005relationship}. A similar guiding principle may be applied to metastable electronic phases. However, this principle is not straightforward because crystallization differs from metal--insulator transitions or magnetic phase transitions in terms of the dynamics of the microscopic components. Atomic diffusion often divergently slows at low temperatures according to the Vogel--Fischer law \cite{tanaka2003possible}, resulting in metastabilization of the supercooled liquid. In contrast, in metal--insulator transitions and magnetic phase transitions, the internal degrees of freedom of the electrons can vary to minimize the free energy of the electronic system even at low temperatures. Therefore, a thermally quenched electronic phase may not be metastable.

\begin{figure*}
\includegraphics[width=131mm]{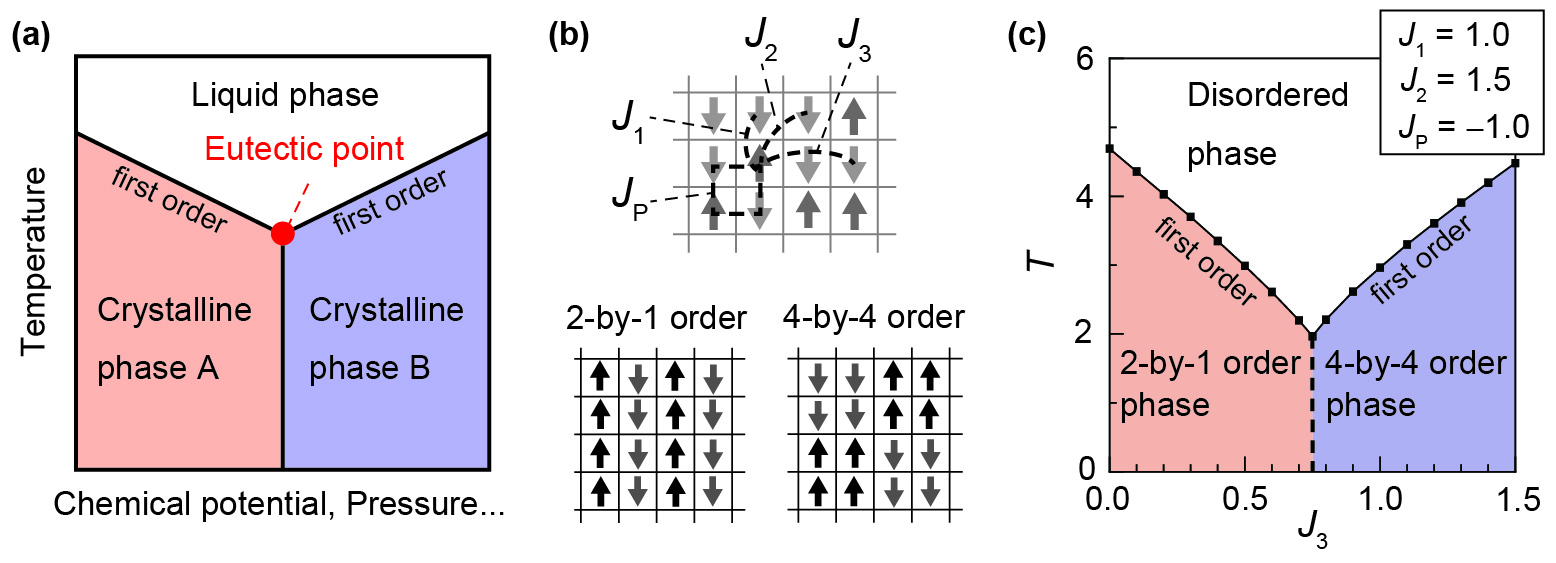}
\caption{\label{Fig1}
(a) Typical eutectic phase diagram. (b) Schematic diagram of the interactions between spins in the Ising model and the ordered phases of interest. (c) $T$--$J_3$ phase diagram where $J_1$, $J_2$ and $J_p$ are fixed to 1.0, 1.5 and -1.0, respectively. Since almost all crystallization processes are first-order phase transitions, the parameters were set so that the ordered phase would also be formed by a first-order phase transition (see Appendix A). A eutectic-like triple point exists at parameters of ($T$, $J_3$) = (2.0, 0.75).
}

\end{figure*}

The present study aims to generalize the dynamics and metastable states near a eutectic-like triple point. The Ising model is suitable for this purpose because it is a typical statistical mechanics model for describing phase transitions involving rearrangement of atoms \cite{lee1952statistical, van1973order} and the internal degrees of freedom of electrons \cite{newell1953theory, castellani1979new}. When eutectic crystallization is modeled with the Ising model, one element is mapped to the up spin and another to the down spin. The conservation of the number of atoms corresponds to the conservation of magnetization, which is defined by the difference in the numbers of up and down spins. This conservation law imposes constraints on changes in the spin configuration, resulting in energy barriers in the process corresponding to the exchange of atoms and the movement of an atom into a vacancy \cite{fratzl1994kinetics, fratzl2000coarsening}. In contrast to eutectic crystallization, the conservation laws are sometimes broken in correlated electron systems, such as in the phase transition from paramagnetic to ferromagnetic phases. Therefore, in the present study, our focus is on the nonconserved Ising model, in which the spin of each site can change independently. The metastability in the phase transition between ordered phases has been studied in the pioneering works \cite{rikvold1994metastable, rikvold2016equilibrium}, but the metastability of the disordered phase has not been clarified. Nevertheless, we find that a thermally quenched disordered phase exhibits metastability near the eutectic-like triple point in the Ising model with competing interactions that favor different ordered structures. This result suggests that the design principles for glass are also applicable to correlated electron systems.

\section{Simulation method}

To implement a eutectic-like triple point in the phase diagram of the Ising model, we consider a two-dimensional square lattice and introduce the Hamiltonian with several types of interactions following a previous study \cite{landau1985phase}:
\begin{equation}
H = J_1\sum_{n.n}\sigma_i\sigma_j +J_2\sum_{2nd}\sigma_i\sigma_j+J_3\sum_{3rd}\sigma_i\sigma_j+J_p\sum_{ring}\sigma_i\sigma_j\sigma_k\sigma_l,
\end{equation}
where $J_1$, $J_2$, $J_3$, and $J_p$ are the nearest neighbor, next-nearest neighbor, third-nearest neighbor and four-body ring interactions, respectively, as shown in Fig.~1(b). $\sigma_i$ represents a spin, which indicates the local configuration at lattice site i and takes a value of +1 or -1. $J_1$ favors a diagonal antiferroic order, and $J_2$ favors a 2-by-1 or 4-by-4 order, depending on the value of $J_3$. $J_p$ represents the many-body spin interactions arising from higher-order effects of electron itinerancy \cite{thouless1965exchange, misguich1999spin} and contributes to the ordering temperature and discontinuity of the phase transition. When the interaction parameters ($J_1$, $J_2$, $J_p$) are fixed to (1.0, 1.5, -1.0), the most stable ordered phase changes from a 2-by-1 order to a 4-by-4 order with variation of $J_3$, as shown in Fig.~1(c). The disordered, 2-by-1, and 4-by-4 phases meet at the parameters of ($T$, $J_3$) = (2.0, 0.75), which correspond to the eutectic-like triple point. Thus, we successfully reproduced the eutectic point with a simple Hamiltonian.

The phase transition kinetics are then simulated at each point of the $T$--$J_3$ phase diagram. The $\sigma_i$ values in an initial state are randomly set at 1 or -1 to simulate a state immediately after thermal quenching from an infinite temperature with an infinite cooling rate. The time evolution of $\sigma_i$ is simulated by the single spin flip and heat bath methods, in which the values of $\sigma_i$ are updated after a Monte Carlo step by the probabilities as

\begin{equation}
P(\sigma_i=1) = \frac{\rm{exp}(-\it{H}(\sigma_i=\rm{1})/\it{T})}{Z},
\end{equation}
\begin{equation}
P(\sigma_i = -1) = \frac{\rm{exp}(-\it{H}(\sigma_i = -\rm{1})/\it{T})}{Z},
\end{equation}
\begin{equation}
Z = \rm{exp}(-\it{H}(\sigma_i=\rm{1})/\it{T}) + \rm{exp}(-\it{H}(\sigma_i=-\rm{1})/\it{T}).
\end{equation}
During one Monte Carlo step, all the sites are updated in a random order. In this study, a Monte Carlo step was considered a unit of time, as in previous studies \cite{bray1994theory, rikvold2016equilibrium, naskar2021metastable}. We imposed periodic boundaries and performed simulations and analyses. In the main text, we present the results for a system size $N$ of $64^2$. The system size dependence was checked, as shown in Appendix B.

\begin{figure*}
\includegraphics[width=131mm]{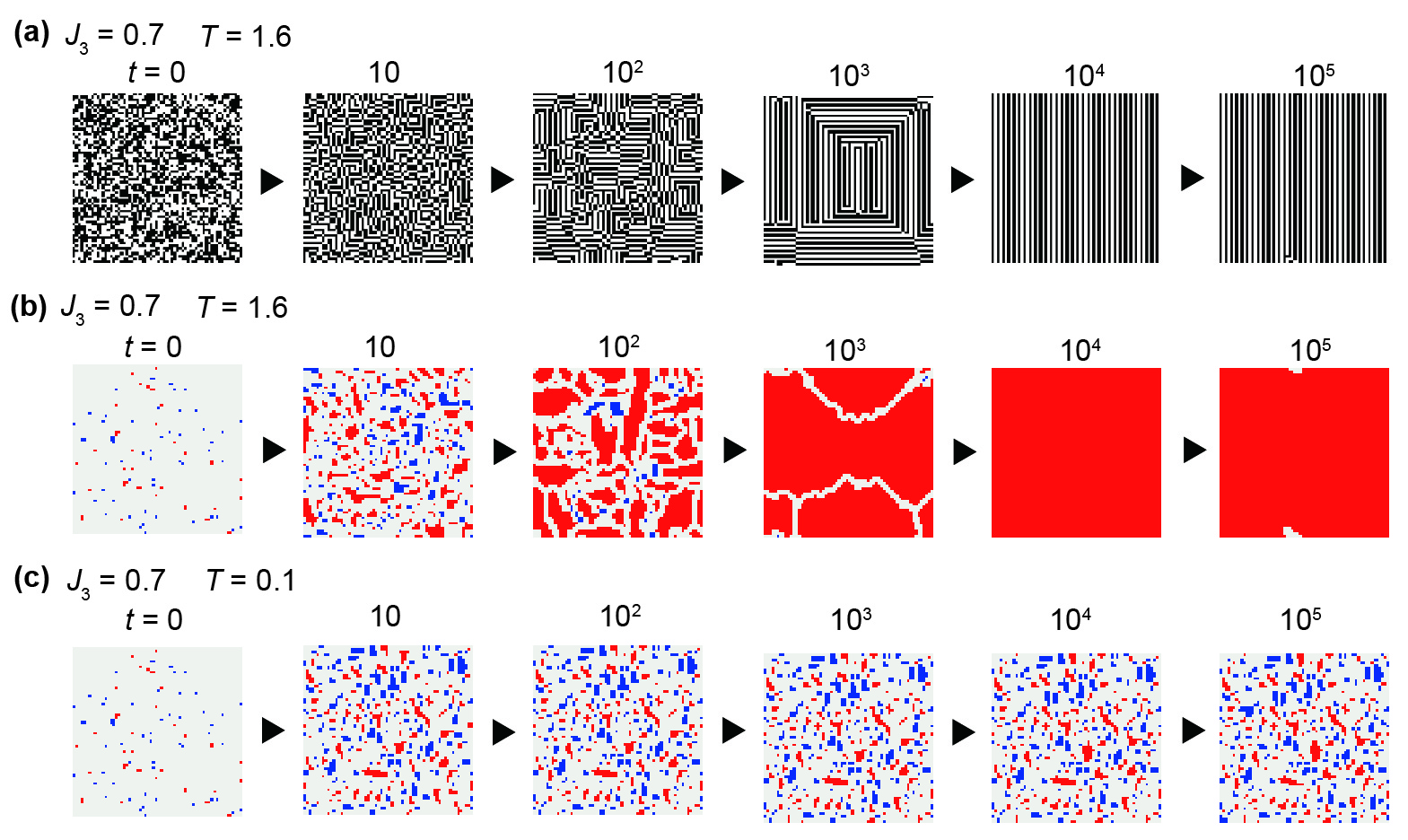}
\caption{\label{Fig2}
(a) Isothermal time evolution of the spin configuration at parameters of ($T$, $J_3$) = (1.6, 0.7). The system size $N$ is $64^2$. (b) Three-colored representation of the time evolution shown in (a). (c) Three-colored representation of the time evolution at parameters of ($T$, $J_3$) = (0.1, 0.7). The red and blue areas represent 2-by-1 and 4-by-4 orders, respectively, and the white area represents all other configurations. The method of converting the spin configurations to three colors is described in the main text.
}
\end{figure*}

\section{Results and Discussions}
\subsection{Emergence of metastability}

To examine metastability, the isothermal time evolution is studied for the parameters of ($T$, $J_3$) = (1.6, 0.7) and (0.1, 0.7), where $J_3$ is slightly decreased from the critical value of the eutectic point, 0.75, so that the 2-by-1 order becomes the most stable phase. At a temperature of $T$ = 1.6, stripe-structured domains representing the 2-by-1 order appear, and the domains gradually increase in size, becoming a uniform 2-by-1 order as time evolves (Fig.~2(a)). To make the phase transition process with time easier to understand, the three types of local structures (2-by-1 order, 4-by-4 order, and disordered) are colored differently according to the following procedure. First, a site is selected, and 3-by-3 sites containing the selected site at the center are extracted. The selected site is colored red if the 3-by-3 sites match the 2-by-1 order, colored blue if they match the 4-by-4 order, or colored white if they do not match either of these orders. Thus, many small 2-by-1 and 4-by-4-order domains appear in the initial stage, and the 2-by-1 domains coarsen over time until they become comparable to the system size (Fig.~2(b)). The ordering kinetics from the supercooled disordered phase presented below were obtained through this procedure.

In contrast to that at $T$ = 1.6, the real space configuration at $T$ = 0.1 is approximately independent of time from $t$ = 10 to $t$ = $10^5$ (Fig.~2(c)). This time-independent behavior means that the disordered phase is stable on the time scale of the simulation. This disordered phase is considered metastable because the free-energetically most stable phase is a uniform 2-by-1 order. The comparison of the configurations at $t$ = 0 and $t$ = $10^5$ indicates that the metastable phase is not a random state of the initial state but a fine domain structure of 2-by-1 and 4-by-4 orders. The formation of a fine structure suggests that local energy optimization occurs in the early stages but that global optimization has not yet occurred.

\subsection{Time-Temperature-Transformation (TTT) diagram}

\begin{figure*}
\includegraphics[width=156mm]{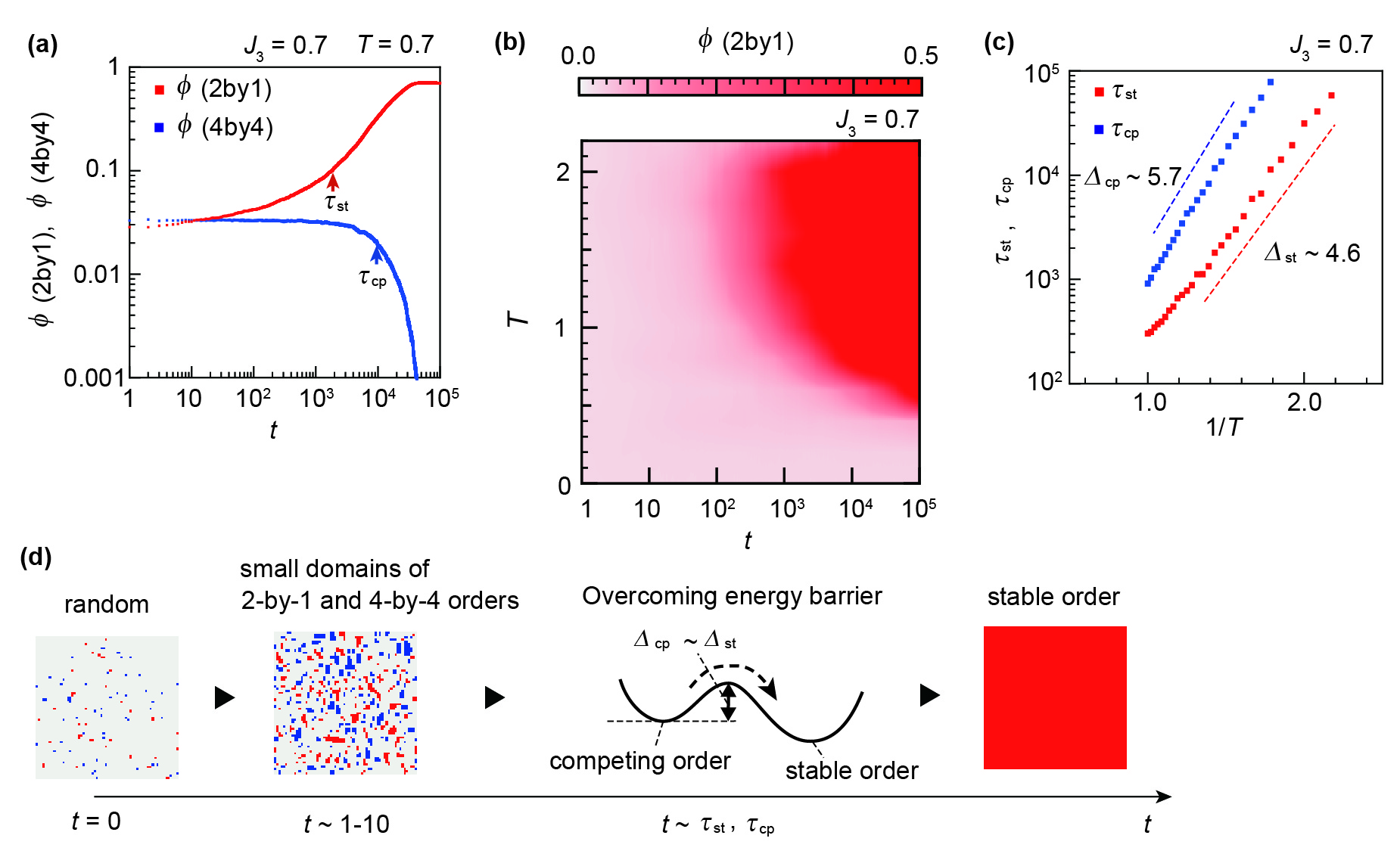}
\caption{\label{Fig3}
(a) Isothermal time evolution of order parameters $\phi$(2by1) and $\phi$(4by4) at parameters of ($T$, $J_3$) = (0.7, 0.7). (b) TTT diagram of the Ising model at a $J_3$ of 0.7. (c) Arrhenius plot of the growth time of the stable 2-by-1 order $\tau_{st}$ and the destruction time of the competing 4-by-4 order $\tau_{cp}$. (d) Schematic representation of the time evolution. From the initial random arrangement, fine 2-by-1 and 4-by-4 domains rapidly form, and when the process of overcoming the energy barrier $\Delta_{st}$, which is comparable to $\Delta_{cp}$, is activated by thermal fluctuations, a uniform 2-by-1 order is formed.
}

\end{figure*}

To obtain an overview of the temperature dependence, we examine the TTT diagram at $J_3$ = 0.7, as thermally quenched metastable phases are often discussed using the TTT diagram \cite{uhlmann1972kinetic, sato2017electronic, sasaki2017crystallization}, which summarizes the temperature-dependent isothermal time evolution of an order parameter. To characterize the extent to which the ordering progresses, we define the order parameters for the 2-by-1 and 4-by-4 phases as follows:
\begin{equation}
\phi(2\mathrm{by}1) = \frac{1}{N} \sqrt{\sum_{\bm{q}=q_x,q_y}\left|\sum_{j=1}^N\sigma_je^{i\bm{q}\cdot r_j}\right|^2},
\end{equation}
\begin{equation}
\phi(4\mathrm{by}4) = \frac{1}{N} \sqrt{\sum_{\bm{q}=q_p,q_m,-q_p,-q_m}\left|\sum_{j=1}^N\sigma_je^{i\bm{q}\cdot r_j}\right|^2},
\end{equation}
where $N$ is the number of sites and $q_x$, $q_y$, $q_p$, and $q_m$ are ($\pi$, 0), (0, $\pi$), ($\pi$/2, $\pi$/2), and ($\pi$/2, $-\pi$/2), respectively. These definitions lead to $\phi$(2by1) and $\phi$(4by4) being equal to one in the uniformly ordered states and approximately zero in an initial state with a random configuration. The isothermal time evolution for several temperatures is obtained by averaging 64 individual runs, as shown in Fig.~3(a). By making a contour map of the time evolution of $\phi(2by1)$ up to $t = 10^5$ for $T$ = 0.001, 0.2, 0.4, 0.6, 0.8, 1, 1.2, 1.4, 1.6, 1.8, 2.0, and 2.2, a TTT diagram is obtained, as shown in Fig.~3(b).

As shown in Fig.~3(b), $\phi$(2by1) rapidly develops over time within a temperature range of $1.5 < T < 2.0$, whereas its evolution dynamics slow at higher and lower temperatures. In particular, within a temperature range of $T < 0.4$, the ordering hardly progresses, and the system does not reach the most stable 2-by-1 phase, at least within the timescale of the simulation, indicating the emergence of metastability. This nonmonotonic temperature dependence is commonly observed in supercooled liquids and arises from two competing temperature dependencies: the energy barriers in the phase transition pathway are less likely to be overcome at low temperatures, whereas the free-energy driving force for ordering becomes greater at low temperatures. The TTT diagram for the present Ising model suggests the existence of an energy barrier.

To characterize the time scale over which a stable ordered phase is formed, we define $\tau_{st}$ as the time at which the order parameter of the stable order exceeds 0.1. The examination of $\tau_{st}$ at various temperatures reveals that $\tau_{st}$ follows an Arrhenius-type behavior (Fig.~3(c)), indicating the presence of a thermal activation barrier $\Delta_{st}$ (Fig.~3(d)). For supercooled liquids, the diffusion of atoms is widely believed to contribute to the energy barrier \cite{fratzl1994kinetics, fratzl2000coarsening}. However, note that a finite $\Delta_{st}$ appears, although individual degrees of freedom are variable without a diffusion process in the present nonconserved system.

To gain insight into the energy barrier, the dynamics of the disappearing competing order are characterized by the time $\tau_{cp}$ at which $\phi$(4by4) is half of the value at $t$ = 1. The examination of $\tau_{cp}$ at various temperatures reveals that $\tau_{cp}$ also shows an Arrhenius-type temperature dependence with a thermal activation barrier $\Delta_{cp}$ (Fig.~3(c)). A possible reason for the close values of $\Delta_{st}$ and $\Delta_{cp}$ is that the rate-limiting process of the phase transition is breaking of competing orders.

\begin{figure*}
\includegraphics[width=182mm]{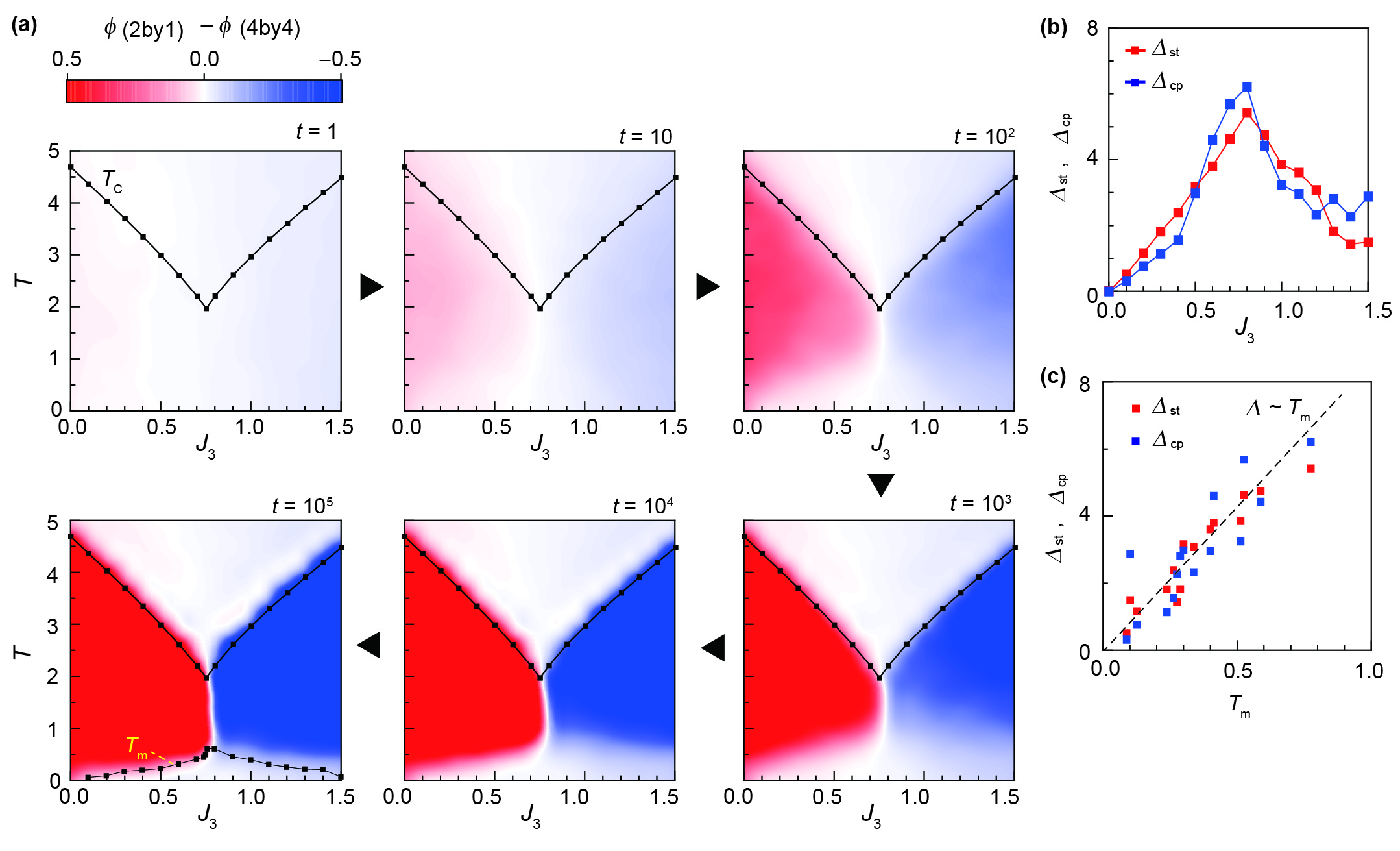}
\caption{\label{Fig4}
(a) Contour plot of the time evolution of the order parameter in the $J_3$ range of 0--1.5. $\phi$(2by1) -- $\phi$(4by4) is plotted, as the formation of the 2-by-1 order is indicated by a positive value (red) and the formation of the 4-by-4 order is indicated by a negative value (blue). $T_m$ is the temperature below which $\phi$(2by1) or $\phi$(4by4) is less than 0.1 at $t = 10^5$. (b) $J_3$ dependence of the energy barriers $\Delta_{st}$ and $\Delta_{cp}$ estimated with the Arrhenius plot shown in Fig.~3(c). (c) Correlations between $T_m$ and the energy barriers $\Delta_{st}$ and $\Delta_{cp}$. A guide to the eye is used to show the proportional relationship.
}

\end{figure*}

\subsection{Energetic competition and metastability}

Because $J_3$ controls the relative stability of the 2-by-1 and 4-by-4 orders, we then examine the $J_3$ dependence of the phase transition kinetics and metastability. To overview the $J_3$ dependence, the values of $\phi$(2by1) -- $\phi$(4by4) are plotted in Fig.~4(a) as contour maps in the $J_3$--$T$ plane at $t$ = 1, 10, $10^2$, $10^3$, $10^4$, $10^5$. As the 2-by-1 (4-by-4) order grows, the red (blue) color becomes clearer in the contour maps (Fig.~4(a)). At a $J_3$ close to 0 or 1.5, the ordering appreciably progresses at $t$ = 10 and is almost complete at $t = 10^3$. As $J_3$ becomes closer to that at the eutectic-like point ($J_3 = 0.75$), both $\phi$(2by1) and $\phi$(4by4) remain at values less than 0.1 even at $t = 10^5$ within a temperature range of $T < 0.5$.

To characterize the region in which the metastable phases exist in the $J_3$--$T$ phase diagram, we define the upper temperature limit of metastability $T_m$ as the temperature below which $\phi$(2by1) and $\phi$(4by4) are less than 0.1 at $t = 10^5$, and the $J_3$ dependence of $T_m$ is also plotted in the $J_3$--$T$ phase diagram at $t = 10^5$ (Fig. 4(a)). $T_m$ is found to take a maximum value near $J_3 = 0.75$, indicating that the system tends to be metastable near the phase boundary between the 2-by-1 and 4-by-4 phases. The energy barriers $\Delta_{st}$ and $\Delta_{cp}$ also take a maximum near $J_3 = 0.75$ (Fig.~4(b)) and are positively correlated with $T_m$ (Fig.~4(c)). Thus, the $J_3$-dependent energy barrier is found to play a key role in controlling the metastability in our model system with a eutectic-like triple point.

\begin{figure*}
\includegraphics[width=162mm]{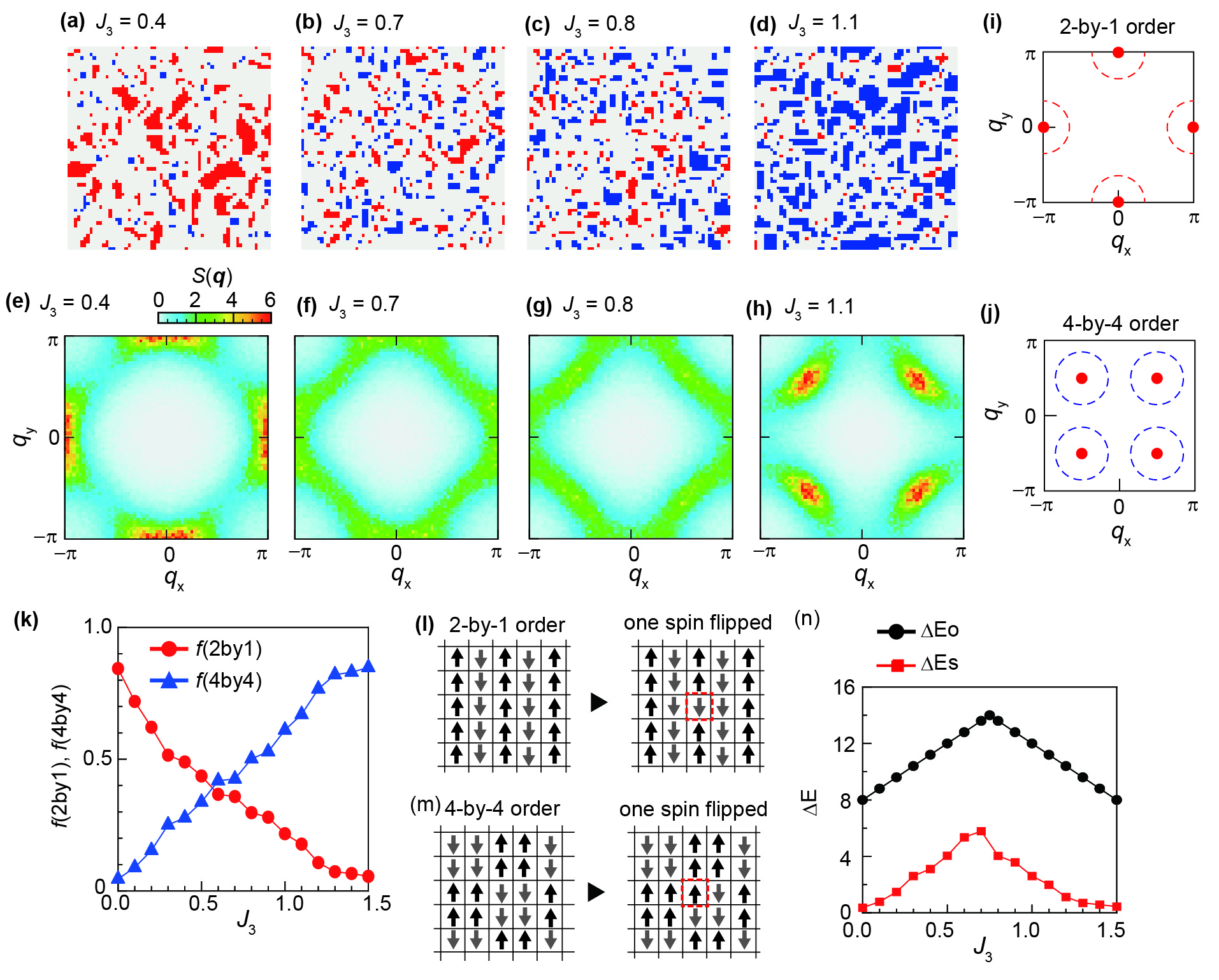}
\caption{\label{Fig5}
(a-d) Three-colored representation of the spin configurations at $T$ = 0.1 for $J_3$ = 0.4 (a), 0.7 (b), 0.8 (c), and 1.1 (d). (e-h) Structure factor $S(\bm{q})$ of the metastable phase shown in (a-d). (i, j) Structure factor $S(\bm{q})$ of the uniform 2-by-1 order (i) and 4-by-4 order (j). (k) $J_3$ dependence of the estimated domain fractions of the 2-by-1 and 4-by-4 orders, $f(2\mathrm{by}1)$ and $f(4\mathrm{by}4)$. See the main text and Appendix C for details of the estimation method. (l, m) Single spin flip process for the uniform 2-by-1 order (l) and 4-by-4 order (m). (n) $J_3$ dependence of the energy barriers $\Delta E_\mathrm{o}$ and $\Delta E_\mathrm{s}$. $\Delta E_\mathrm{o}$ is the energy increase due to the single spin flip processes shown in (l) for $0.75 < J_3 < 1.5$ and (m) for $0 < J_3 < 0.75$. The scaled energy barrier $\Delta E_s$ equals $f(4\mathrm{by}4)\Delta E_\mathrm{o}$ for $0<J_3<0.75$ and $f(2\mathrm{by}1)\Delta E_\mathrm{o}$ for $0.75<J_3<1.5$.
}

\end{figure*}

The good correspondence between the evolution of the energy barriers and that of the metastability suggests that the local configuration is correlated with the metastability. To address this issue in more detail, below, we examine the disordered metastable state in terms of the structure factor. At a temperature of $T$ = 0.1 and $J_3$ values of 0.4, 0.7, 0.8, and 1.1, the metastable disordered phases contain both 2-by-1 and 4-by-4 domains, the fractions of which depend on $J_3$ (Fig.~5(a)-(d)). To quantify the domain fractions, the real-space configurations are converted into the structure factor $S$($\bm{q}$) as
\begin{equation}
S(\bm{q}) = \frac{1}{N} \left|\sum_{j=1}^N\sigma_je^{i\bm{q} \cdot \bm{r_j}}\right|^2,
\end{equation}
as shown in Fig.~5(e)-(h). The intensities of $S$($\bm{q}$) near ($\pm\pi$, 0) and (0, $\pm\pi$) correspond to the fraction of the 2-by-1 domains (Fig.~5(i)), and those near ($\pm\pi$/2, $\pm\pi$/2) correspond to the fraction of the 4-by-4 domains (Fig.~5(j)). The domain fractions of the 2-by-1 and 4-by-4 orders, $f(2\mathrm{by}1)$ and $f(4\mathrm{by}4)$, are estimated by integrating $S$($\bm{q}$) over the circular areas near these $\bm{q}$ values, as shown in Fig.~5(i) and (j). Whereas $\phi$(2by1) and $\phi$(4by4) reflect $S$($\bm{q}$) only for the $q$ values corresponding to the uniformly ordered structure, $f(2\mathrm{by}1)$ and $f(4\mathrm{by}4)$ integrate $S$($\bm{q}$) around the $q$ values. Because the Fourier transformation of a small domain includes the $S$($\bm{q}$) around the $q$ values, $f(2\mathrm{by}1)$ and $f(4\mathrm{by}4)$ are considered more suitable for evaluating the fractions of 2-by-1 and 4-by-4 domains in the fine domain structure. $f(2\mathrm{by}1)$ and $f(4\mathrm{by}4)$ are found to vary monotonically as a function of $J_3$ (Fig.~5(k)), reflecting the $J_3$ dependence of the domain structures, as shown in Fig.~5(a)-(d).

Having clarified the microscopic structure of the metastable phase, we discuss how the energy barrier found in the simulation can be evaluated on the basis of the microscopic structure. We consider a single spin flip to break the uniform competing order, which corresponds to Fig.~5(l) for $J_3$ = 0.75--1.5 and to Fig.~5(m) for $J_3$ = 0--0.75. The energy barrier $\Delta E_\mathrm{o}$ required for these processes is calculated and plotted in Fig.~5(n) and is found to be much larger than $\Delta_{st}$. This discrepancy in the energy barrier reflects the fact that the single spin flips in the uniformly ordered phase are not a direct factor determining the robustness of the metastability. Because the metastable phases that we are discussing are highly disordered and the locally formed order is often surrounded by disordered states, the discrepancy between $\Delta E_\mathrm{o}$ and $\Delta_{st}$ is rather reasonable. To account for the probability that a locally formed competing order is surrounded by disordered states, the energy barrier is scaled by the domain fraction as $\Delta E_s$ = $f(4\mathrm{by}4)\Delta E_\mathrm{o}$ for $0 < J_3 < 0.75$ and $f(2\mathrm{by}1)\Delta E_\mathrm{o}$ for $0.75 < J_3 < 1.5$. We then obtain absolute values similar to those of $\Delta_{st}$ (Fig.~5(n)). The correspondence between the scaled energy barrier $\Delta E_s$ and that obtained by the Arrhenius law $\Delta_{st}$ suggests a method for quantitatively estimating energy barriers from microscopic structures.

\subsection{Discussions}

In the mean-field approximation, the metastable solution of a disordered phase is lost when the system is supercooled to below the spinodal point \cite{miyashita2022collapse}. Similarly, in the present Ising model at $J$ = 0.7, the local minimum close to $\phi$(2by1) = 0 is lost at $T$ = 2.18, which is only a few percent below $T_c$ (see Fig.~6(c) in the appendix). However, the disappearance of a metastable solution does not necessarily indicate the absence of a practically stable phase. Indeed, the present study shows that a practically metastable phase exists even when the system is deeply supercooled. This type of metastability is often observed in solid solutions produced by rapid cooling. For example, in the TiO$_2$--VO$_2$ system, spinodal decomposition occurs at temperatures below 830 K, but under rapid cooling, a stable solid solution with a uniform distribution of Ti and V ions can be obtained at room temperature \cite{hiroi2013spinodal}, meaning that spinodal decomposition is kinetically arrested. Such kinetic arrest of spinodal decomposition is widely accepted in the context of conserved systems. In nonconserved systems, whether kinetic arrest occurs is nontrivial because the most stable phase has a uniform order and there is no thermodynamic driving force for decomposition. The appearance of a metastable phase in the present Ising model demonstrates that kinetic arrest of spinodal growth can also occur in a nonconserved system, implying the universality of kinetic arrest beyond obvious free energy minima.

Finally, we discuss the possible relevance of the present simulations for correlated electron materials. Transition metal oxides are typical examples of materials with eutectic-type geometries in electronic phase diagrams \cite{tokura1999colossal, tokura2006critical, shibuya2010metal}. In manganites with a composition in which two electronic phases compete, a glassy electronic state appears under the cooling process with a standard cooling rate due to the effect of randomness \cite{tokura2006critical}. This sensitivity to randomness may indicate a tendency for a metastable disordered phase to appear under thermal quenching. In terms of the energetic competition between multiple ordered phases, geometrical frustration, as in the case of triangular lattices, can also be effective in realizing metastable phases. For example, in quarter-filled triangular lattice systems $\theta$-(BEDT-TTF)$_2$X, several charge-ordered patterns are energetically close to each other, indicating energetic competition \cite{seo2000charge, pramudya2011nearly, mahmoudian2015glassy, rademaker2018suppressed}. Metastable charge glass phases are less likely to appear under a cooling process with a certain cooling rate when the triangular lattice is distorted \cite{sato2014systematic, oike2015phase}, indicating that frustration is effective for inducing metastable phases. Thus, frustrated systems, which have attracted much attention for the exploration of exotic quantum phases, are promising for the realization of metastable electronic phases.

\section{Concluding remarks}
The present study aims to generalize the principle of metastability established in metallurgy with the goal of applying it to correlated electron systems. We have demonstrated that a thermally quenched disordered phase exhibits metastability near the eutectic-like triple point in the Ising model with competing interactions. This simple model of metastability enables us to discuss how metastability varies with the interaction strength and how it correlates with local structures. The nature of metastability revealed by the model is likely relevant to experimentally observed metastable electronic phases, suggesting a link to correlated electron systems.

Microscopic simulations address phenomena of a limited time range due to computational constraints. Therefore, the appearance of a metastable phase may simply indicate that the time range is insufficient to reach the most stable phase. In addition, the physics research of correlated electron systems has focused mainly on their most stable phases. Thus, numerically found metastable phases have not received much attention. However, many metastable materials have been created by exploring states before the most stable state is reached. If we focus on the physics of metastability, then a numerically obtained metastable phase would become a valuable research target.

\appendix
\section{Appendices}
\section{APPENDIX A: Determination of the phase diagram}
The phase transition temperature of the Ising model was determined from the temperature dependence of the order parameter when the free energy reaches a minimum value (Fig.~6). When $J_3$ is less than 0.75, a 2-by-1 order forms at the phase transition temperature, whereas when it is greater than 0.75, a 4-by-4 order forms. The order parameter shows a discontinuous change at the phase transition temperature, indicating a first-order phase transition. The free energy was calculated using the Wang--Landau algorithm \cite{wang2001determining}.

\begin{figure}
\includegraphics[width=89mm]{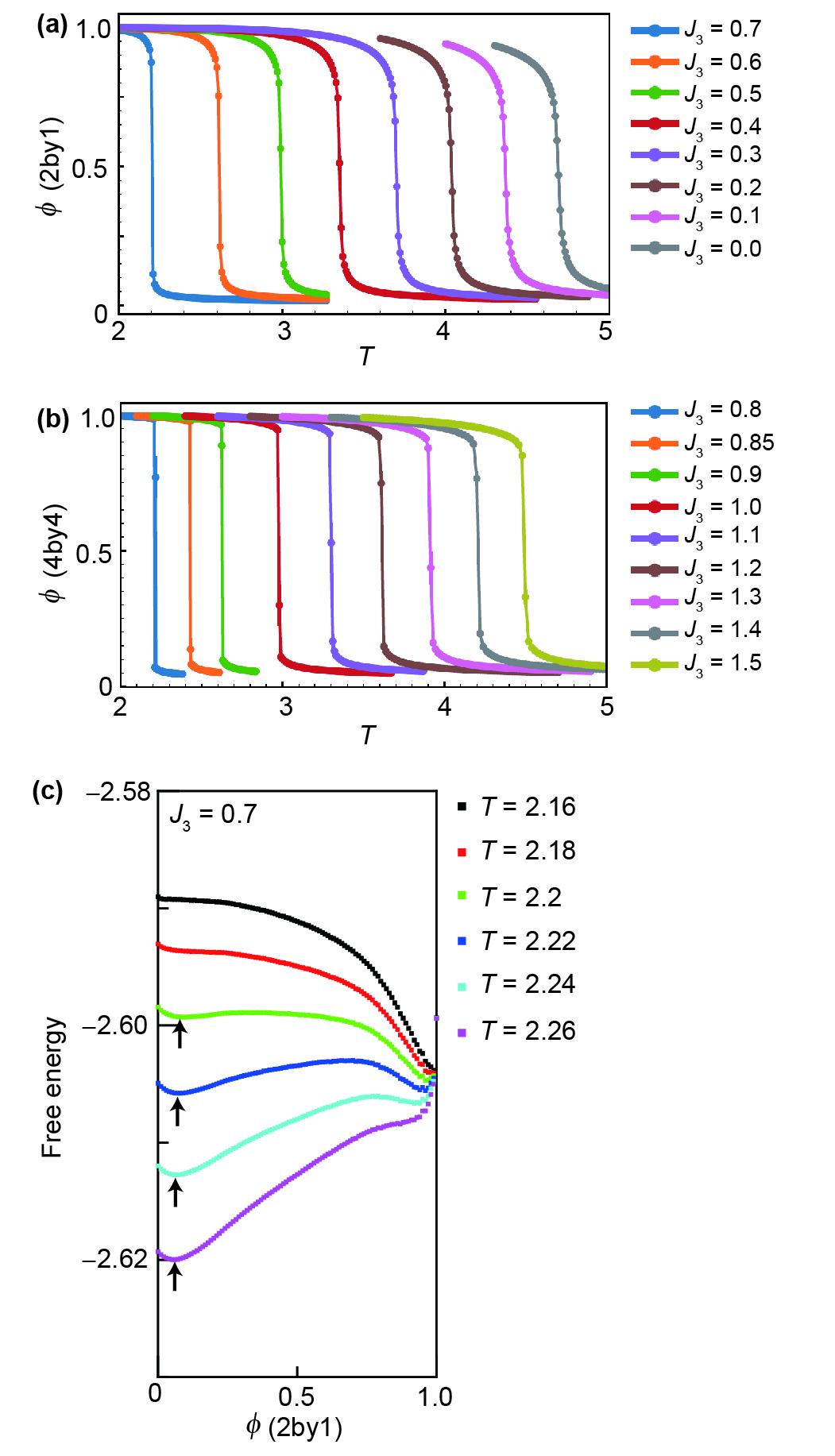}
\caption{\label{Fig6}
(a, b) Temperature dependence of the order parameters $\phi(2by1)$ for $J_3$ = 0.0--0.7 (a) and $\phi(4by4)$ for $J_3$ = 0.8--1.5 (b). (c) Free energy landscape for $J_3$ = 0.7 near the ordering temperature. The arrows indicate the minimum free energy near $\phi$(2by1) = 0.}

\end{figure}

\section{APPENDIX B: System size dependence}
To check the system size dependence, a comparison between system sizes $N (= L^2)$ of $64^2$ and $128^2$ is shown in Figure 7. Although the typical domain size is not dependent on the system size, the order parameter is by definition dependent on the system size for the following reasons. The order parameter $\phi(2by1)$ can be rewritten as
\begin{equation}
\phi(2by1) =\sqrt{\frac{1}{N}\left(\sum_{\bm{q}=q_x,q_y}S(\bm{q})\right)},
\end{equation}
and the structure factor $S$($\bm{q}$) can be rewritten as
\begin{equation}
S(\bm{q}) = \frac{1}{N}\sum_{j,k=1}^N<\sigma_j\sigma_k>e^{i\bm{q} \cdot (\bm{r_j}-\bm{r_k})},
\end{equation}
where $<\sigma_j\sigma_k>$ is the correlation function of spins at sites $j$ and $k$. $\sigma_j\sigma_ke^{i\bm{q} \cdot (\bm{r_j}-\bm{r_k})}$ is 1 if sites $i$ and $j$ are in the same domain of the ordered structure with wavenumber $\bm{q}$ and can otherwise take a random value whose absolute value is one. Therefore, the value obtained by taking the sum of this quantity over $k$ is approximately equal to $N_q^j$, which is the domain size of the ordered structure with wavenumber $\bm{q}$ containing site $j$. The structural factor can then be expressed approximately as
\begin{equation}
S(\bm{q}) \sim \frac{1}{N}\sum_{j=1}^N<N_q^j>.
\end{equation}
The domain size and the number of domains are approximately proportional to $\xi^2$ and $N/\xi^2$, respectively, where $\xi$ is the correlation length. The structure factor can then be approximated by taking the sum over $j$ as
\begin{equation}
S(\bm{q}) \sim \xi^2.
\end{equation}
The order parameter is then expressed as
\begin{equation}
\phi(2by1) \sim \frac{\xi}{L}.
\end{equation}
Thus, $\phi(2by1)$ depends on the system size when $\xi$ does not. We observed that $L\phi(2by1)$ remains independent of the system size up to the time scale at which $L\phi(2by1) \sim \xi$ approaches $L$ (Fig.~8). When the typical domain size is equal to the system size, the order estimate described above is no longer valid. Therefore, this scaling does not hold when $\phi(2by1)$ is close to 1. In the present study, we focus on the dynamics when $\phi(2by1)$ is less than 0.5 to discuss the properties that are independent of the system size. In this $\phi(2by1)$ range, $L\phi(2by1)$ shows similar behavior for $L$ = 64, 128, 256, and 512, as seen in the TTT diagrams (Fig.~9), so we examined the $T$--$J_3$ dependence in detail for $L$ =64.

\begin{figure}
\includegraphics[width=89mm]{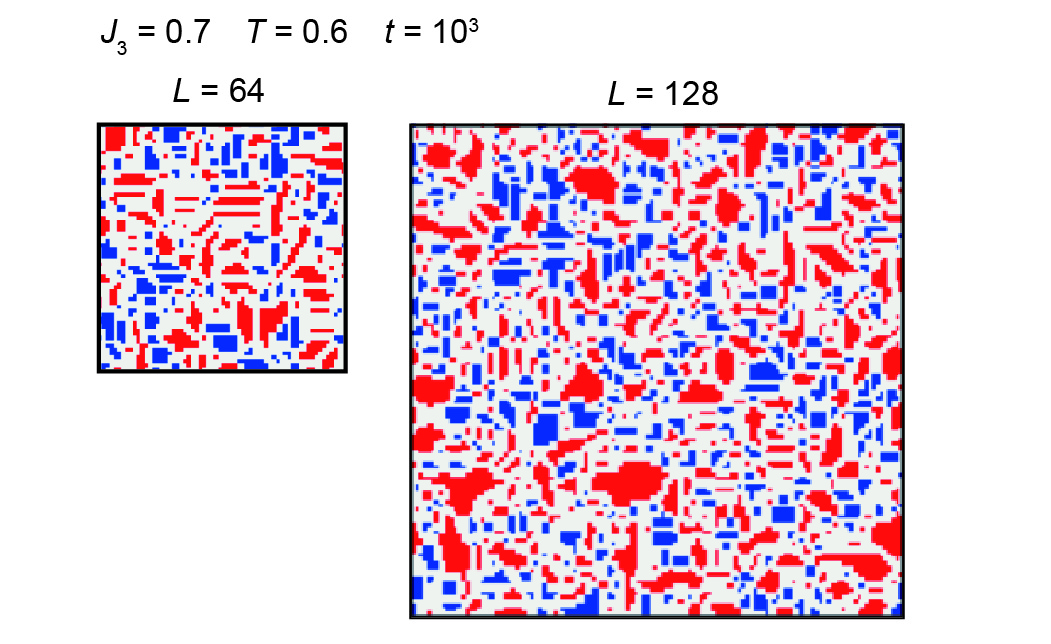}
\caption{\label{Fig7}
Spin configurations at parameters of ($T$, $J_3$) = (0.6, 0.7), $t = 10^3$ and $L = 64$ and $128$.}

\end{figure}

\begin{figure}
\includegraphics[width=89mm]{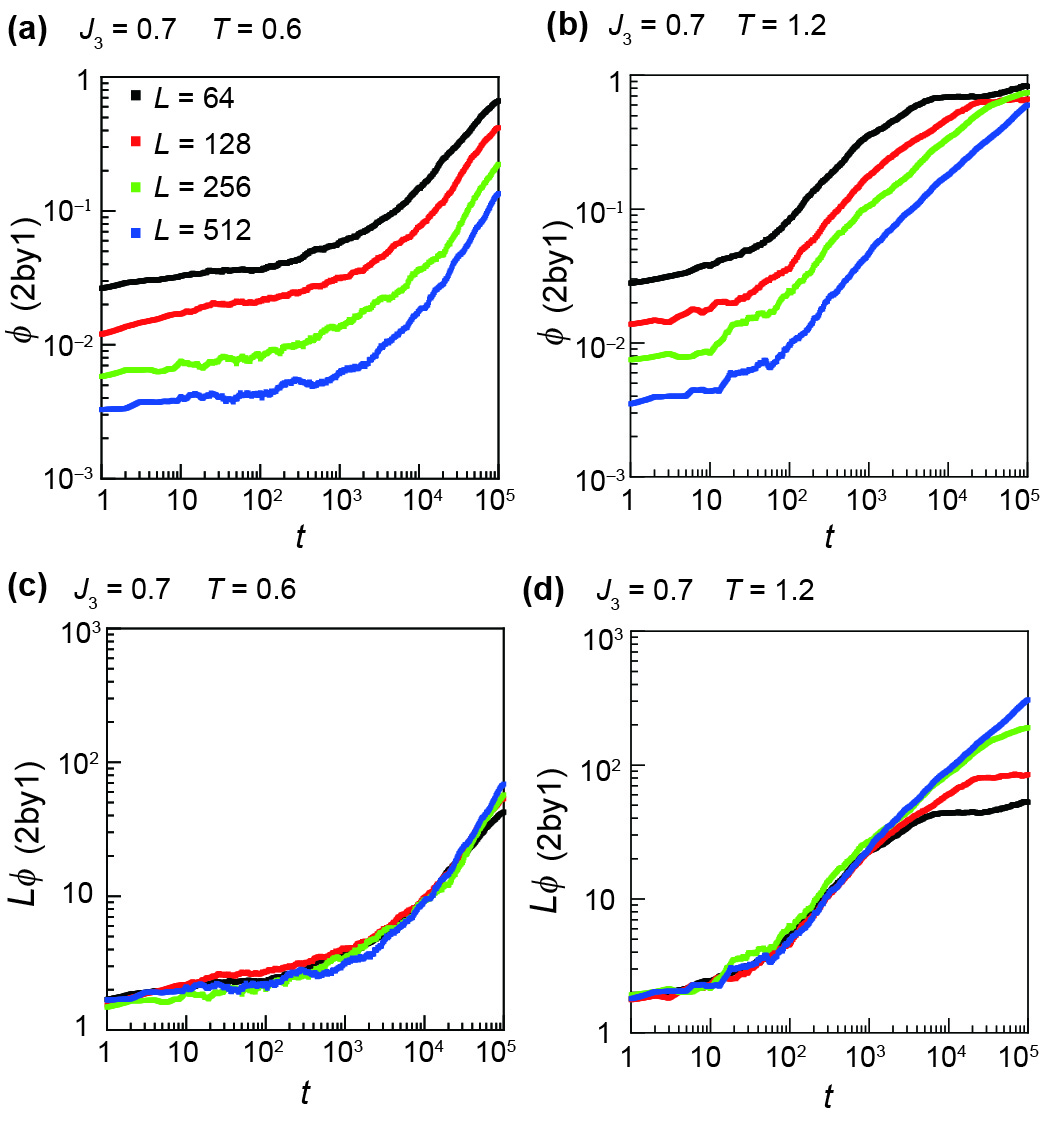}
\caption{\label{Fig8}
(a, b) Isothermal time evolution of $\phi(2by1)$ for ($T$, $J_3$) = (0.6, 0.7) (a) and ($T$, $J_3$) = (0.6, 0.7) (b). (c, d) Isothermal time evolution of $L\phi(2by1)$ for ($T$, $J_3$) = (0.6, 0.7) (c) and ($T$, $J_3$) = (0.6, 0.7) (d). The size dependence was examined at $L$ = 64, 128, 256 and 512. }

\end{figure}

\begin{figure}
\includegraphics[width=89mm]{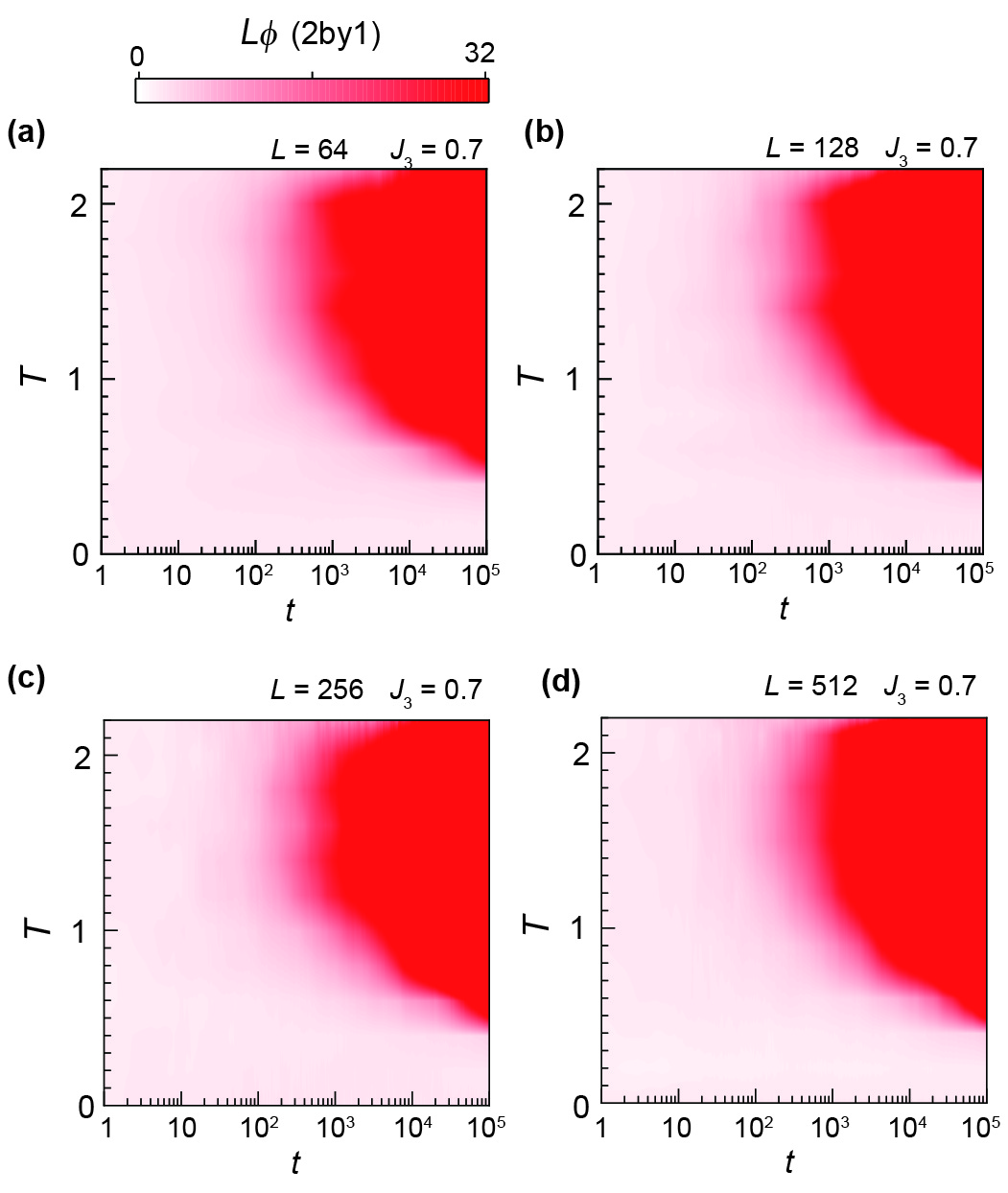}
\caption{\label{Fig9}
(a-d) TTT diagram plotted based on $L\phi(2by1)$ for $J_3$ = 0.7 and $L$ = 64 (a), 128 (b), 256 (c) and 512 (d). }

\end{figure}

\section{APPENDIX C: Estimation of the domain fractions}
The fractions of domains of 2-by-1 and 4-by-4 orders in the disordered phase were estimated from the structure factor as follows. In a uniform 2-by-1 ordered state, $S(\bm{q})$ is equal to $N$ if $\bm{q}$ is either ($\pi$, 0) or (0, $\pi$), which are the points indicated in Fig.~4(i), and zero if $\bm{q}$ is any other value. In a uniform 4-by-4 ordered state, $S(\bm{q})$ is equal to $N$ if $\bm{q}$ is ($\pi$/2, $\pi$/2), ($\pi$/2, $-\pi$/2), ($-\pi$/2, $\pi$/2), or ($-\pi$/2, $-\pi$/2), which are the points indicated in Fig.~4(j), and zero if $\bm{q}$ is any other value. When the 2-by-1 and 4-by-4 orders form a domain structure, $S(\bm{q})$ takes a finite value around the values of $\bm{q}$ corresponding to the 2-by-1 and 4-by-4 orders. The values obtained by integrating the $S(\bm{q})$ of these diffuse spots over the $\bm{q}$ range shown in Fig.~4(i) and (j) indicates how much areas the 2-by-1 and 4-by-4 domains occupy. Therefore, by normalizing these values by $N$, the domain fractions of the 2-by-1 and 4-by-4 orders are estimated as
\begin{equation}
f(2by1) = \frac{1}{N}\sum_{q \sim q_x,q_y}S(\bm{q}),
\end{equation}
\begin{equation}
f(4by4) = \frac{1}{N}\sum_{q \sim q_p,q_m,-q_p,-q_m}S(\bm{q}).
\end{equation}
\begin{acknowledgments}
\section*{ACKNOWLEDGMENTS}
H.O. thanks H. Ohtani, M. Enoki, K. Nakanishi and H. Kageyama for fruitful discussions. This work was supported by JST PRESTO (Grant No. JPMJPR21Q2), the Center of Innovation for Sustainable Quantum AI (SQAI), JST (Grant No. JPMJPF2221), JST CREST (Grant No. JPMJCR24I1), and JSPS KAKENHI (Grant Nos. 22H01164, 23H04861, 22K03508 and 24H01609). MANA is supported by World Premier International Research Center Initiative (WPI), MEXT, Japan. 
\end{acknowledgments}


%

\end{document}